\pdfoutput=1

\documentclass[11pt]{article}

\usepackage[final]{acl}

\usepackage{times}
\usepackage{latexsym}

\usepackage[T1]{fontenc}

\usepackage[utf8]{inputenc}

\usepackage{microtype}

\usepackage{inconsolata}

\usepackage{graphicx}

\usepackage{xcolor}
\usepackage{tabularray}
\usepackage{comment}

\title{Tackling Social Bias against the Poor:\\ A Dataset and Taxonomy on Aporophobia}

\author{
    Georgina Curto\textsuperscript{\rm 1}, Svetlana Kiritchenko\textsuperscript{\rm 2}, Muhammad Hammad Fahim Siddiqui\textsuperscript{\rm 3},\\{\bf Isar Nejadgholi\textsuperscript{\rm 2}, Kathleen C. Fraser\textsuperscript{\rm 2}}\\
    \textsuperscript{\rm 1}United Nations University Institute in Macau, Macau SAR, China\\
    \textsuperscript{\rm 2}National Research Council Canada, Ottawa, Canada\\
    \textsuperscript{\rm 3}University of Ottawa, Ottawa, Canada\\
   \small \textbf{Correspondence:} \texttt{curto@unu.edu, svetlana.kiritchenko@nrc-cnrc.gc.ca}
}

\begin{document}
\maketitle

\begin{abstract}
{\textit{\textbf{Content Warning:} This paper presents textual excerpts that may be offensive or upsetting.}}

Eradicating poverty is the first goal in the United Nations Sustainable Development Goals. However, 
\textit{aporophobia}---the societal bias against people living in poverty---constitutes a major obstacle to designing, approving and implementing poverty-mitigation policies. This work presents an initial step towards operationalizing the concept of aporophobia to identify and track harmful beliefs and discriminative actions against poor people on social media. In close collaboration with non-profits and governmental organizations, we conduct data collection and exploration. Then we manually annotate a corpus of English tweets from five world regions for the presence of (1) direct expressions of aporophobia, and (2) statements referring to or criticizing aporophobic views or actions of others, to comprehensively characterize the social media discourse related to bias and discrimination against the poor. Based on the annotated data, we devise a taxonomy of categories of aporophobic attitudes and actions expressed through speech on social media. Finally, we train several classifiers and identify the main challenges for automatic detection of aporophobia in social networks. This work paves the way towards identifying, tracking, and mitigating aporophobic views on social media at scale. 
\end{abstract}

\section{Introduction}

Poverty is a multidimensional phenomenon that affects 712 million people worldwide. Sub-Saharan Africa continues to be the region with the highest number of people (411.15 million) living under the poverty line of \$2.15 a day \cite{TheWorldBank2024}. Yet, poverty is not only a challenge for developing economies. In the United States, there are 37.9 million people living in poverty  \cite{Creamer2022}, and in Europe, 4.6 million people are at risk of poverty \cite{Eurostat2023}. While the traditional efforts to mitigate poverty have been losing effectiveness in the last decades \cite{WorldBank2022}, the capability approach to human development \cite{Sen2001,Nussbaum2012,Comim2018} has become the main interdisciplinary alternative to the traditional economic frameworks on poverty mitigation.

The capability approach defines poverty as one's lack of capabilities (where \textit{capabilities} are the 
things that people can be and do if they choose so, like being in good health, getting married, being educated, or traveling) to conduct a meaningful life with dignity \cite{Sen1979}. Therefore, the focus of the approach is on increasing individuals' well-being and agency. While classical utilitarian approaches center on the redistribution of wealth \cite{Bentham2010,Mill2017}, these often constitute a palliative strategy that fails to address the underlying dimensions of poverty \cite{Sen1979}. In turn, it has been argued that the rhetoric of equal opportunity \cite{Rawls1971} could contribute to blaming the poor for their condition \cite{Fishkin2016,Sandel2020a}. This is especially relevant in countries such as the United States and Canada, which embrace the narrative of the ``land of opportunity'' \cite{Desmond2023, curto2024}, but in reality have low rates of social mobility \cite{Chetty2014}.

This paper aligns with the capability approach to human development and aims to provide language resources and new insights to tackle poverty by identifying, tracking and analyzing societal bias against people living in poverty, known as \textit{aporophobia}. Aporophobia has been defined as the ``rejection, aversion, fear and contempt for the poor'' \cite{Cortina2022}. It increases the burden of poverty, impacting the well-being of this vulnerable group, and constitutes an obstacle to poverty mitigation. When the poor are blamed for their situation and considered undeserving of help, it is harder for policymakers to approve and implement poverty-mitigation strategies 
\cite{Arneson1997,Everatt2009,Nunn2009}. 

The way society imagines and acts towards the poor is part of the poverty phenomenon, independently of how one decides to measure it \cite{Sherman2001}. Aporophobia is rooted in our beliefs and manifests through different degrees of attitudes and actions \cite{Comim2020}. In this study, we identify and characterize aporophobia by analyzing how it is expressed through language, leveraging natural language processing (NLP) techniques. 
We distinguish between \textit{direct} expressions of personal aporophobic views and \textit{reporting} on or discussing aporophobic views and actions of other people or institutions. While both types of expressions signify aporophobic bias in the society, this distinction can be informative for countering measures. 
`Direct' expressions of aporophobia may directly be confronted through removal from public view (for severe expressions) or addressed with counterspeech. The `reporting' instances, on the other hand, inform more complex mitigation strategies, often outside the scope of online platforms.

Since there are no available data resources annotated for aporophobia, we start by collecting and annotating English texts from various regions that refer to poor people. This provides us with a better understanding of the diversity of commonly expressed beliefs and behaviors regarding the poor. We focus on social media, and in particular X (formerly Twitter), as our data source for this paper.\footnote{We note that while some aporophobic posts can constitute hate speech according to social media platforms' policies on safe content, most aporophobic texts do not violate such safety guidelines.} While the discourse on this platform does not comprehensively represent the opinions of the world population, it constitutes a valuable resource for exploring vast amounts of data and conducting a preliminary analysis that can be complemented with other sources in further studies. 

We analyze the data and create a taxonomy of aporophobic actions expressed through speech as the first resource to characterize aporophobia on social media.  Then, we experiment with machine learning models to evaluate the viability of automatic aporophobia detection. The results of this evaluation are relevant for future work aimed at identifying and tracking aporophobia at scale, both regionally and globally. This would allow us to study the potential correlation between aporophobia and socio-economic indicators as well as the impact of various policies  in different world regions. 

Thus, the main contributions of this work are as follows: 
\begin{itemize}
    \item We present a novel data collection and annotation method, which relies on unsupervised topic modeling, to considerably increase the proportion of target (aporophobic) instances in a data sample to be labeled. We ensure fair representation of various geographical regions in the dataset by oversampling posts from the underrepresented regions, including Africa, South Asia, and Oceania.  
    \item Grounded in cognitive science, philosophy of discrimination, and human development literature, as well as the inputs from specialized non-profits and governmental bodies, we manually annotate social media posts and devise a taxonomy of expressed aporophobic behaviors corresponding to various degrees of discriminative action. These categories are grouped under two types of speech: `Direct' (expressing the speaker's own aporophobic views) and `Reporting' (stating or criticizing the views and behaviors of others). 
    \item We release the first-ever dataset, DRAX (Direct and Reported Aporophobia on X), manually annotated for aporophobia and containing 1,816 English tweets from five world regions.\footnote{The dataset is publicly available for research purposes: \url{https://svkir.com/projects/aporophobia-data.html}}
    \item Using the annotated data, we investigate the feasibility of automatic aporophobia detection with various language models. Then, through an error analysis, we identify future directions for improving aporophobia detection models. 
\end{itemize}

\section{Related Work}
In the last decades, an important body of work has been devoted to studying and mitigating discrimination against vulnerable groups, namely those defined
on the grounds of sex, race, religion, political or other opinion, national or social origin, and other characteristics \cite{UnitedNations1966,CouncilofEurope2010}. Sadly, aporophobia has not received the attention it deserves in the literature as a distinct field of study \cite{Curto2022}. It was not until the 1990s that the term was coined by philosopher Adela Cortina \cite{Cortina2022}. The study of aporophobia unveils the prejudices on poverty, which has an impact on essential topics, including public policy \cite{Comim2020}, the judicial system \cite{TerradillosBasoco2020} and the economic power dynamics \cite{Desmond2023}. More importantly, aporophobia affects the dignity of the disadvantaged as ``the mere fact of being poor is itself cause for being isolated, left out, looked down upon, alienated, pushed aside, and ignored by those who are better off.'' \cite{Narayan2002}.   

Interdisciplinary research in AI has attended to the detection and mitigation of biases with two ultimate goals: \textit{ethical AI}, which aims to mitigate inherent biases in models \cite{Curto2022,Blodgett2020,Lalor2022} and \textit{AI for social impact}, which constitutes policy-relevant research where AI is used as a tool to detect and mitigate societal biases \cite{aguilera2024can,Havaldar2024,Li2024}. Our work distinctly focuses on the latter. 

AI researchers, and specifically NLP scholars, have addressed social biases online by creating annotated datasets \cite{poletto2021resources}, developing algorithms to detect biased language  \cite{alkomah2022literature, jahan2023systematic} and to counter it \cite{fraser2023makes, nejadgholi2024challenging,hassan2023discgen}, extending these models to accommodate multiple languages \cite{chhabra2023literature} and cultural contexts \cite{lee2023hate, lee2024exploring}, and establishing ethical guidelines for deploying such technologies \cite{kiritchenko2021confronting,garg2023handling}. However, most of these attempts focused on racism and sexism \cite{mansur2023twitter} and, in some cases, bias against immigrants \cite{pitropakis2020monitoring}, sexual minorities \cite{chakravarthi2022overview} or religious groups \cite{vidgen2020detecting}. 

Only recently has aporophobia emerged as a topic in NLP research that revealed the pervasive yet under-studied issue of bias against poor people on social media \cite{Kiritchenko2023}. Research studies explored the association between crime and poverty on social media across eight geographically diverse countries \cite{curto2024}, the correlation between aporophobia and increased wealth inequality \cite{aguilera2024can}, negative public attitudes towards people experiencing homelessness \cite{ranjit2024oath}, and the impact of aporophobia on the stigmatization of specific groups \cite{brate2024bayesian}. While these studies solidified the necessity of confronting aporophobia using AI tools, our work is the first step in providing the necessary resources for training models to detect and track this form of bias at scale. 

\begin{figure*}[t!]
\centering
  \includegraphics[width=0.95\textwidth]{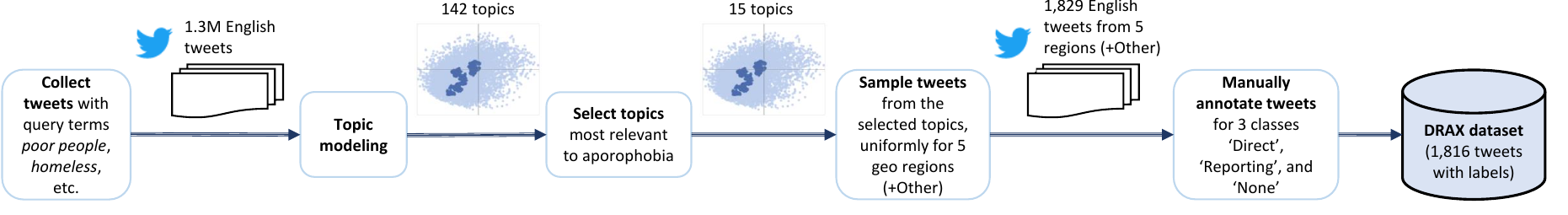}
  \caption{The process diagram for the novel methodology to collect and annotate data.}
  \label{fig:flowchart}
\end{figure*}

Creating resources to detect social biases online poses significant challenges \cite{yin2021towards,kovacs2021challenges}. Firstly, the identification of biases needs to be rooted in cognitive science \cite{Allport1954,Kahneman2011}, the philosophy of discrimination \cite{Young2022,Honneth1996,Taylor2009} and sociology \cite{Fuchs2018,Kozlowski2019}, among other disciplines, and involve domain experts. 
Secondly, while biases have been traditionally studied from a single-dimension perspective, they should be considered in a multi-axial approach \cite{Hoffmann2019}, 
for example, examining how aporophobia and xenophobia aggravate one another \cite{Cortina2022}.  

Thirdly, collecting data to identify social biases is complex due to its scattered distribution on social media and the subtle, complex ways such beliefs are expressed \cite{vidgen2020directions}. Random sampling results in a very small number of examples of the harmful class, which is insufficient for training classifiers \cite{vidgen2019much}. Researchers have tackled this issue by employing targeted search techniques, using keywords and hashtags, and also boosted random sampling techniques \cite{madukwe2020data,naseem2021survey}, 
which has been shown to introduce biases and over-reliance on keywords in trained classifiers \cite{wiegand-etal-2019-detection}. 
In this work, we first use generic keywords to collect data on the target population and then employ unsupervised topic modeling, which takes into account context beyond mere keyword matching, allowing us to concentrate on annotating data from the most relevant topics. 

Finally, annotating such data is a demanding task, requiring precise definitions to capture nuanced expressions of harmful language and must address annotators' disagreements, stemming from subjective interpretations across different cultures and personal experiences \cite{vidgen2020directions, curry2024subjective}.  While many studies have applied a binary classification, few works have attempted to create more detailed fine-grained categories \cite{salminen2018anatomy,kirkSemEval2023}. We annotate our dataset for three high-level categories and then devise a fine-grained taxonomy that is both grounded in the literature and on the findings from the dataset annotation. 

\section{DRAX Dataset Collection and Annotation}

\subsection{Tweet Collection and Data Sampling}

Through the Twitter API,\footnote{Data was collected before the introduction of the paywall.} we collected tweets in English between 25 August 2022 and 23 November 2022 using the following query terms: \textit{the poor} (used as a noun as opposed to an adjective, as in `the poor performance'), \textit{poor people}, \textit{poor ppl}, \textit{poor folks}, \textit{poor families}, \textit{homeless}, \textit{on welfare}, \textit{welfare recipients}, \textit{low-income}, \textit{underprivileged}, \textit{disadvantaged}, \textit{lower class}. (Further details on query term selection and tweet pre-processing are in Appendix~\ref{sec:appendix_tweets}.) 
By using tweet location (‘place’ field), where available, or user location field, we grouped tweets into the following six regions: North America, Europe, Africa, South Asia, Oceania, and Other.\footnote{'Other' group includes tweets with no geographical information in either field, tweets with location strings that could not be parsed into a valid location, and tweets originated from countries in other world regions, for which only very limited English data was available (e.g., Middle East).}  
Most of the collected tweets (over 62\%) fall into the Other category; 26\% come from North America, about 7\% are from Europe, and only about 1\% were written in Africa, South Asia, and Oceania.    

Annotating a random sample of tweets for the presence of aporophobia would be inefficient since only a small percentage of tweets express aporophobic attitudes. On the other hand, sampling data instances that contain specific (harmful) keywords can significantly restrict and bias the collected data. Therefore, we propose a novel data collection method, based on topic modeling (depicted in Figure~\ref{fig:flowchart}). Previous work showed that topic modeling is an effective tool in uncovering the subtle details within a raw dataset, taking into account the full context  \cite{nejadgholi2020cross,bourgeade-etal-2023-learn,piot2024metahate}. 

First, we masked the query terms in the retrieved tweets, and applied unsupervised topic modeling, using BERTopic \cite{grootendorst2022bertopic}, on a random sample of 600K tweets (see Appendix~\ref{sec:appendix_bertopic} 
for technical details). Overall, 142 topics were identified. Then, we manually analyzed the topic words and the most representative example tweets from the obtained topics and selected 15 topics highly relevant to the concept of aporophobia. The selected topics include subjects such as blame, crime, substance abuse, immigrants and refugees, and racism, among others. (The full list of topics with example tweets is available in Table~\ref{tab:topics} in Appendix.) 
Finally, tweets from each of the 15 topics were randomly sampled to satisfy the following two conditions: (1) uniform distribution by region (equal amounts of tweets are sampled from each of the six geographical regions), and (2) uniform temporal distribution (equal amounts of tweets are sampled from each of the three months of the collected data). For topics with a broad relevant content (on crime, drug abuse, immigrants and refugees, and racism), around 250 tweets were sampled per topic; for the rest of the topics, around 100 tweets per topic were included. Since prevalence of topics differed across the regions, we could not always sample the exact number of tweets for each topic and region. 
Overall, this process resulted in 1,829 tweets for manual annotation.

\subsection{Data Annotation}
\label{sec:data_annotation}
We manually annotated each of the selected textual instances (tweets) with one of the three categories: (1) `Direct Aporophobia', defined as text expressing the speaker’s own aporophobic views, (2) `Reporting Aporophobia', defined as text stating or criticizing the aporophobic views and behaviors of others, or (3) `None' (none of the above). 
While `Reporting' posts do not necessarily promote aporophobic views, both categories `Direct' and `Reporting' comprise instances referring to aporophobic attitudes or actions and present evidence on levels of aporophobic bias in the society.

In the first round of annotation, a team of two field experts, with multidisciplinary knowledge on human development, poverty studies, aporophobia, and NLP (authors of this paper), annotated 25 tweets from each selected topic and devised the classification schema and detailed annotation guidelines (available in Appendix~\ref{app:guidelines}). Non-profits and governmental organizations specialized in poverty provided guidance during the annotation, and validated the approach and the resulting outcomes in a process that was performed in several iterations.  Then, a third annotator (also an author of this paper) was trained using the guidelines and the annotated sample. Subsequently, all selected tweets were independently annotated by two annotators. 

The overall inter-rater agreement was 71\% (Cohen's kappa of 0.57). Disagreements mostly emerged due to different interpretations of the annotated texts since tweets are short and sometimes lack the necessary context for precise interpretation of the intended meaning. Furthermore, aporophobia, as with any discriminatory phenomenon, is hard to define precisely for borderline cases. After individual annotations were completed, all the disagreements were discussed among the annotators and resolved in the final dataset. Thirteen tweets were judged as lacking sufficient context for annotation and were removed. 

\subsection{Characteristics of the DRAX Dataset}
\label{sec:dataset_properties}

\begin{figure}[t!]
  \includegraphics[width=0.95\columnwidth]{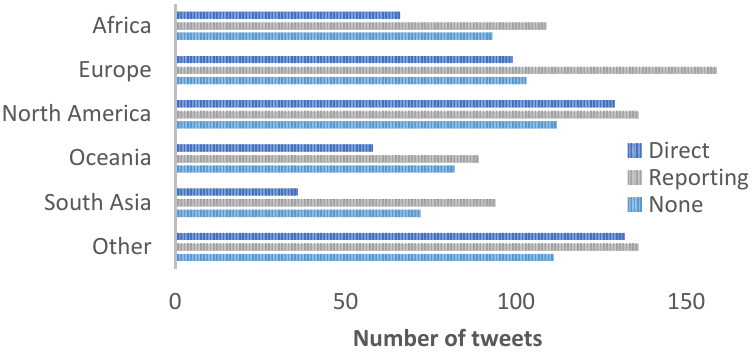}
  \caption{Class distribution in DRAX per geographical region.}
  \label{fig:data-stats-regions}
\end{figure}

The resulting dataset contains 1,816 annotated tweets. We refer to it as DRAX (Direct and Reported Aporophobia on X). 
There are 520 (29\%) instances labeled as `Direct Aporophobia', 723 (40\%) instances labeled as `Reporting Aporophobia', and 573 (32\%) instances labeled as `None'. 

Figure~\ref{fig:data-stats-regions} shows the class distribution per geographical region.\footnote{In Table~\ref{tab:data-stats-countries} in Appendix, we break this distribution down per country. Not all countries in a region are represented equally. For most regions, the overwhelming majority of tweets originate from the largest English-speaking country, such as the U.S. for North America and the U.K. for Europe.}  
For each region, we annotated 202--377 tweets. 
One can observe that tweets from North America and those grouped as `Other' in terms of location, contain the highest proportion of `Direct' aporophobic actions expressed through speech ($\sim$35\%).\footnote{We note a very similar class distribution for the tweets from North America and those grouped as `Other'. Looking at the content of tweets in the `Other' group, we conjecture that a large portion of tweets with no identifiable geo-location originate from the U.S.}  
In contrast, subsets from South Asia, Africa, and Oceania include the lowest proportion of tweets in the `Direct' category (18\%, 25\%, 25\%, respectively). 
The data from Europe presents the highest amount of tweets `Reporting' aporophobic actions of others. 

The topics with the highest proportion of `Direct' aporophobic statements are those referring to drug addiction and mental health issues, and immigrants and refugees (see Figure~\ref{fig:data-stats-topics} and Table~\ref{tab:topics} in Appendix). Other topics with a high proportion of tweets in the `Direct' category refer to crime, homeless encampments, smell, alcohol addiction, and fear.  Such messages often stereotype poor people, and especially the homeless group, as substance addicts and criminals, or express the general attitudes of fear and contempt toward the group. Another topic contains texts communicating the views of rejecting immigrants and refugees since they do not bring any resources and depend on the state's support. 

Topics with a high rate of `Reporting Aporophobia' refer to racism, crime, hatred, the military, laws and courts, laws and regulations, and blaming the poor. Messages in these topics often criticize the governments and people in power for taking advantage and discriminating against poor people through unfair enforcement of laws and regulations, and blaming all social and economic issues on the lower socio-economic class. Black communities are seen as the most targeted since race-based discrimination results in both social and economic disadvantages.

\section{Taxonomy of Aporophobic Actions Expressed Through Speech}

\begin{figure*}[t!]
  \includegraphics[width=0.95\textwidth]{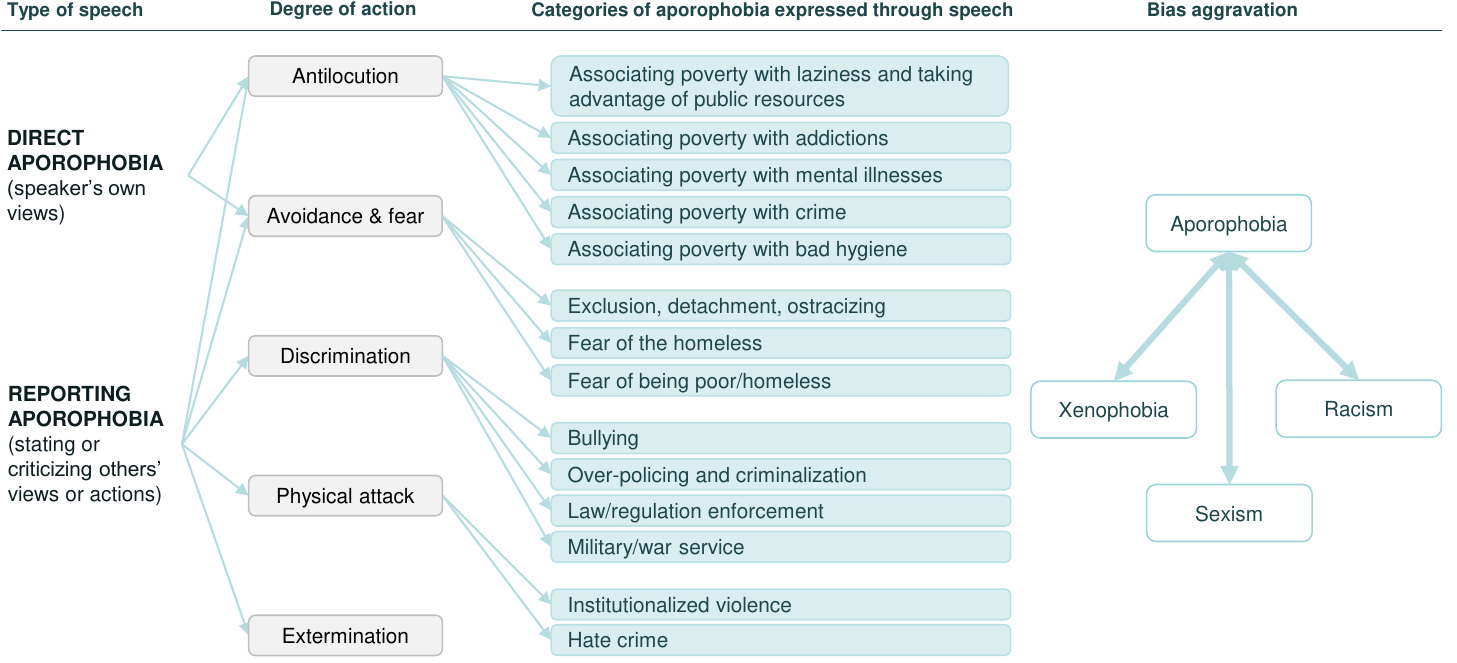}
  \caption{Taxonomy of categories for three levels of classification of aporophobia: Type of speech, Degree of action, Categories of aporophobia expressed through speech.}
  \label{fig:taxonomy}
\end{figure*}

Based on the qualitative analysis of the DRAX dataset, the conceptual framework for the nature of prejudices, bias and discrimination \cite{Allport1954,Kahneman2011,Taylor2009,Honneth1996,Blodgett2020,Fuchs2018} and the concept of aporophobia \cite{Cortina2022,Comim2020}, we devise a taxonomy of aporophobic actions expressed through speech (depicted in Figure~\ref{fig:taxonomy}). 

\vspace{5mm}
\textbf{The first level of classification ``Type of Speech''} 
(Figure~\ref{fig:taxonomy}, left column) defines whether the text constitutes `Direct' aporophobia or whether it is `Reporting' the aporophobic views or actions of others, as defined in Section~\ref{sec:data_annotation}. Here, note that statements refuting stereotypes, such as ``\textit{most poor people aren't violent}'' are also included in the `Reporting' category. Despite the fact that these sentences negate the stereotypes, they at the same time acknowledge and may indirectly reinforce them \cite{Beukeboom2019}.  
Both `Direct' and `Reporting' categories reveal the levels of aporophobic bias in the society. However, these two types of expressions may require different mitigating strategies. 

\textbf{The second level of classification ``Degree of Action''} (Figure~\ref{fig:taxonomy}, second left column) corresponds to Allport's categorization for the different degrees of negative action resulting from prejudices: antilocution (i.e., negative verbal remarks), avoidance and fear, discrimination, physical attack, and extermination \cite{Allport1954}. It allows classifying aporophobic instances based on their severity: from `Antilocution' in sentences such as ``\textit{poor people tend to abuse alcohol}'' to the most severe category of `Extermination', evidenced in sentences such as ``\textit{poor people should be euthanized}''. 
We consider that the texts are a form of speech, and as such can \textit{directly} express a person’s aporophobic attitudes through antilocution or fear (e.g., statements expressing prejudice) or  \textit{report} on aporophobic attitudes and behaviors of others. However, behavior such as discrimination, physical attack, and extermination are actions that occur out in the real world, and therefore can only be \textit{reported} on social media. 
For that reason, the former two categories (antilocution and fear) appear under both direct and reported aporophobia, but the latter three categories (discrimination, physical attack, and extermination) only appear under the reporting category.

\textbf{The third level of classification ``Categories of Aporophobia Expressed Through Speech''} (Figure~\ref{fig:taxonomy}, middle column) corresponds to the different themes of aporophobia expressed through language, based on the analysis of the DRAX dataset. 
We see themes on negative attitudes towards poor people (antilocution) due to the stereotypical associations of this group with substance abuse, mental illnesses, crime, bad hygiene, and the general belief that poverty is the result of laziness and lack of agency to hold one's life in control (e.g., `\textit{he had so many chances but he told me that he would rather do drugs and be homeless}''). Another common theme is fear and ostracism of poor people and especially people in a situation of homelessness (e.g., ``\textit{homeless encampments are taking over streets and parks, residents are in desperation for the unsafe conditions}''). Many posts report on the tendency of people in power to blame any negative circumstances on poor people (bullying) as in the statement ``\textit{It's not easy to blame politicians and their friends. It's easier to blame the poor and their families.}''  Furthermore, criminalization of poverty and over-policing as well as disproportionate law and regulation enforcement on people lacking resources have been actively criticized in posts from across the world (e.g., ``\textit{Typical, the law is meant only for the poor.}'', `` \textit{American justice system: rich people get warnings and probation, poor people go to prison.}''). We also see messages reporting incidents of physically attacking homeless persons (e.g., ``\textit{Two arrested for beating and looting a homeless old man}''). 
Appendix~\ref{app:guidelines} provides additional examples for each of the categories. 

It is important to emphasize that most common themes associated with poverty (addictions, crime, laziness) put the blame for living in poverty on the poor themselves. However, in reality, the overwhelming majority of the poor population in North America (the region with the highest representation in the dataset and the highest level of `Direct' aporophobic instances in DRAX) are either born into poverty or became poor due to circumstances beyond their control, such as disabilities, divorce, illness, old age, or low wages \cite{UnitedNations2018,Desmond2023}.

\textbf{Bias Aggravation:} The taxonomy (Figure~\ref{fig:taxonomy}, on the right) also highlights the often intersectional nature of bias, when aporophobia can be aggravated by as well as affect other types of discrimination. 
We particularly found numerous examples where xenophobia and racism are linked with aporophobia.
For example, a recurrent argument that appears on social media is the ``need to take care of our homeless first'' as an argument to reject migrants and ethnic minority groups (e.g., ``\textit{We don't want them! Can't help our homeless, but sure let's put a roof over migrants heads}'').
Another example is tweets from the United States that reveal how the poor are often assumed to be people of color and immigrants (e.g., ``\textit{only Black people are on welfare}'').  These preliminary findings unveil existing prejudices, since in fact  44\% of the population in poverty in the U.S. are white, followed by 28.4\% Hispanic, of any race \cite{Shrider2023}. 
Correlations between sexism and aporophobia also constitute an interesting area of study in NLP. 
Social science literature shows that non-market care work is often excluded from mainstream economics analysis, which can marginalize women and undervalue their (unpaid) contributions to the community \cite{Folbre2021}.

\section{Automatic Aporophobia Detection}

In this section, we investigate the viability of automatic aporophobia detection in text data. For this purpose, we experiment with a range of state-of-the-art NLP techniques, using the DRAX dataset, and establish benchmark results. Further, we analyze the model performance and the different types of errors to point out future research directions for the effective automatic detection of aporophobia.  

We approach the task of automatic aporophobia detection as a three-class classification problem (`Direct', `Reporting', or `None'), and employ pre-trained language models from the BERT family as well as generative large language models (LLMs). Specifically, we leverage BERT \cite{BERT}, BERTweet \cite{Bertweet}, RoBERTa \cite{RoBERTa}, and DistilBERT \cite{DistilBERT}---four of the most prominent BERT-family models known for their high performance on various text classification tasks. We fine-tune these models on DRAX to optimize their ability to identify and categorize aporophobic actions expressed through language. Additionally, we perform zero-shot and few-shot experiments using two open-source LLMs, Llama 3.1 405B Instruct\footnote{\url{https://huggingface.co/meta-llama/Llama-3.1-405B-Instruct}} \cite{dubey2024llama} and Mixtral 8x22B Instruct,\footnote{\url{https://huggingface.co/mistralai/Mixtral-8x22B-Instruct-v0.1}} and three OpenAI's generative LLMs, GPT-3.5 Turbo, GPT-4o, and GPT-4o mini.\footnote{\url{https://platform.openai.com/docs/models}} 
These experiments aim to assess the models' ability to detect aporophobia with minimal or no task-specific training. In Appendix~\ref{app:toxicity}, we also present the results of the experiments using two existing models trained to detect general toxicity and hate speech. These experiments reveal that the existing models, though effective in general toxicity detection, fall short in accurately identifying aporophobic content, thus highlighting the need for specialized models in this domain.

We split the DRAX dataset into a training and a test subset chronologically, using data from the first two months for training and the last month for testing.\footnote{To ensure the reliability of our results, we also performed a 3-fold cross-validation and obtained similar results.} This approach allows us to simulate a real-world scenario where models are trained on past data and deployed on newer, unseen data. 
Table~\ref{tab:data-stats} shows the number of instances in each subset. 

We fine-tune models from the BERT family on the training portion and evaluate the performance on the test subset. For generative LLMs, we crafted 20 distinct prompts, each providing a brief definition of the three classes (`Direct,' `Reporting,' and `None'). The models are instructed to output only one of these labels. For few-shot prompting, we added nine paraphrased examples from the training set, three for each class label. We tested these prompts on a validation set (stratified by region and label),
and selected the best-performing prompt for each model for evaluation on the test set. 
The overall best-performing prompt (few-shot GPT-4o) is listed in Appendix (Table~\ref{tab:best-prompts}). 
For performance evaluation, we employ standard evaluation metrics: accuracy and support-weighted average precision, recall, and F1-score. 
For fine-tuned models, we repeat experiments 3 times with different fixed seeds, and report the average results.\footnote{The results across the runs are stable, each within 1-2\% of the average.}
The results are presented in Table \ref{tab:results}. 

\begin{table}
  \centering
  \small
  \begin{tabular}{lrrrr}
    \hline
    \textbf{Data subset} & \textbf{Direct} & \textbf{Reporting} & \textbf{None} & \textbf{Total} \\
    \hline
    Training set & 347 & 494 & 389 & 1,230\\
    Test set & 173 & 229 & 184 & 586\\
    \hline
    Overall & 520 & 723 & 573 & 1,816\\
    \hline
  \end{tabular}
  \caption{The training and test dataset statistics.}
  \label{tab:data-stats}
\end{table}

Among the tested models, RoBERTa achieved the highest overall performance with an F1-score of 64\%. While not very high, such performance may already be adequate for the task of tracking major changes in societal attitudes towards people living in poverty in response to local or global events, such as wars, natural disasters, or new policies. For more fine-grained analyses, further improvements on classification models are needed.

\setlength{\tabcolsep}{5pt}
\begin{table}
  \centering
  \small
  \begin{tabular}{lrrrr}
\hline
\textbf{Model}  & \textbf{Acc.} & \textbf{P} & \textbf{R} & \textbf{F1}   \\
\hline
\textit{\textbf{Fine-tuned models}} & & & &\\                           $\ \ \ \ $BERT   & 0.62    & 0.62    & 0.62    & 0.61  \\
$\ \ \ \ $BERTweet  & \textbf{0.64}     & 0.63  & \textbf{0.64}   & 0.63  \\
$\ \ \ \ $DistilBERT  & 0.63  & 0.63 & 0.63  & 0.62  \\
$\ \ \ \ $RoBERTa    & \textbf{0.64}    & \textbf{0.64} & \textbf{0.64} & \textbf{0.64}\\[5pt]

\textbf{\textit{Generative LLMs}}    & & & & \\      $\ \ \ \ $Llama 3.1 405B~- zero shot & 0.57  & 0.61      & 0.59   & 0.56 \\
$\ \ \ \ $Mixtral 8x22B~- zero shot & \textbf{0.63}  & \textbf{0.65}      & \textbf{0.61}   & \textbf{0.60} \\
$\ \ \ \ $GPT-3.5 Turbo - zero shot  & 0.49 & 0.49 & 0.50   & 0.49          \\
$\ \ \ \ $GPT-4o mini~- zero shot & 0.53 & 0.53 & 0.53 & 0.53          \\
$\ \ \ \ $GPT-4o~- zero shot & 0.60  & 0.58      & 0.60   & 0.59 \\[3pt]
$\ \ \ \ $Llama 3.1 405B~- few shot & 0.60  & 0.61      & 0.61   & 0.60 \\
$\ \ \ \ $Mixtral 8x22B~- few shot & 0.61     & 0.65      & 0.59   & 0.59 \\
$\ \ \ \ $GPT-3.5 Turbo~- few shot & 0.54 & 0.56  & 0.54   & 0.54          \\
$\ \ \ \ $GPT-4o mini~- few shot  & 0.55   & 0.56  & 0.55  & 0.55          \\
$\ \ \ \ $GPT-4o~- few shot & \textbf{0.63}     & \textbf{0.69}      & \textbf{0.63}   & \textbf{0.63} \\
\hline
  \end{tabular}
\caption{The performance of fine-tuned and generative large language models on the DRAX dataset. The evaluation metrics are overall accuracy (Acc.), and support-weighted average precision (P), recall (R), and F1-score (F1). The highest numbers in each group are in bold.}
\label{tab:results}
\end{table}
\setlength{\tabcolsep}{6pt}

Below, we summarize the key insights gained from manual inspection of the best performing classifier's errors. 

\vspace{1mm}
\noindent\textbf{Highly nuanced contexts:} Classifiers struggle to effectively incorporate the nuances of the context.  Surprisingly, we observe that our best-performing classifier makes the most mistakes in detecting the `None' class, labeling more than 50\% of it as `Direct' or `Reporting' aporophobia. Deeper investigations of errors uncover additional challenges. First, the classifiers struggle to interpret the nuances in semantic associations between entities mentioned in the text and negative traits or behaviors (who is accused of what). For example, ``\textit{the local homeless shelter director and the CVS pharmacist were busted for dealing opioids at the shelter}'' is misclassified as `Direct'. Second, statements that challenge stereotypical assumptions are not always recognized as such and misclassified as `Direct', e.g., 
``\textit{Perception. Homeless are often wrongly assumed to be drug addicts, drinkers, ex-offenders etc.}''  Another type of error has to do with the direction of the association. For example, the belief that all homeless people are drug addicts is aporophobic, but the viewpoint that drug addiction can lead to homelessness is not. Yet, we observe some instances where the model makes such mistakes, for example misclassifying the text ``\textit{Take heroin and you'll find yourself ruined, homeless, fallen in the gutters}'' as `Direct' (instead of `None'). 

\vspace{1mm}
\noindent\textbf{Impact of tone:} The model tends to interpret texts written in a more formal tone as more objective (and classify them as `Reporting') and more informal texts as expressing subjective views (and classify them as `Direct'). For example, the post ``\textit{Very hard to take care of the poor with no policing. No one wants to invest in the poorer areas as they will lose their businesses due to crime.}'' is misclassified as `Reporting' (instead of `Direct'), whereas the post ``\textit{They’re not mutually exclusive Paddy. The govt should be able to house migrants and the homeless.}'' is misclassified as `Direct` (instead of `Reporting').

\vspace{1mm}
\noindent\textbf{Impact of topic bias:} A number of topics are skewed towards one of the classes, leading to higher performance on the majority class for the topic. For example, the model struggles to distinguish between `Direct' aporophobic actions towards refugees and innocuous statements about this group, due to `Direct' aporophobia being the dominant class in the topic that discusses this group (57\%).  For example, the message ``\textit{So sad and contradictory, with so many refugees and homeless people around the world}'' is misclassified as `Direct' aporophobia (instead of `None'). 

\vspace{2mm}
\noindent\textbf{Generalization across geographical regions:} The best performance is achieved on tweets from Oceania (F-score of 68\%) and North America and South Asia (both with F-score of 67\%), and the worst performance is obtained on data from Europe and `Other' groups (F-score of 61\%). 
Further experiments with the training data sampled only from the dominant regions (North America and Other) show a substantial decrease in results for Africa and Oceania, demonstrating the importance of balanced data sampling across various geographical regions (see Appendix~\ref{app:sampling} for the detailed results). 

\vspace{5pt}

Overall, the above observations suggest that specialized language models capable of handling more nuanced contexts may be of help here, yet will substantially increase the computational costs due to further training and utilizing bigger language models. Additionally, larger and more linguistically and semantically diverse training data can mitigate some of the issues. Synthetic data generation techniques can potentially offer the required scale and diversity of the data while reducing the costs of manual data selection and annotation. Finally, machine learning techniques for handling class imbalance (e.g., downsampling the dominant class) may reduce the impact of topic bias. By creating and releasing the annotated dataset, we provide the resources and encourage further work on this challenging NLP task.

\section{Discussion and Future Work} 

This article examines aporophobia expressed through online speech and provides new insights to better understand the multidimensionality of poverty. In particular, we create the first annotated dataset and the first taxonomy for the phenomenon of aporophobia  in social networks. 
These resources present preliminary evidence of how the cultural contexts in different regions affect the beliefs and degrees of negative action towards people living in poverty.
While North America has the highest percentage of tweets ($\sim$35\%) expressing `Direct' aporophobia, posts from European countries express a higher percentage of `Reporting' aporophobic actions. Future research will further explore this geographical comparative, to provide insights on whether the higher levels of `Direct' aporophobia in North America could be aligned with the tradition of meritocracy \cite{Sandel2020a} and the narrative of the ``the American dream''. This line of work has important implications for policy making \cite{UnitedNations2018}, since caricatured narratives towards the poor could reinforce stereotypes and constitute an obstacle for addressing the shortcomings of social protections.

Future work will extend the analysis to cover other languages and data sources, including multi-modal data combining text with images and video. Moreover, based on the existing dataset and taxonomy, further studies can examine aporophobia in other types of written texts, such as legal or corporate texts, bringing new insights on aporophobia in diverse real-world scenarios. In addition, while the labeling in the current dataset includes only the categories of `Direct', `Reporting', and `None', future work can incorporate the annotation of all the categories of aporophobia identified in the taxonomy. Further, analyzing and characterizing aporophobia as an aggravator of other types of bias (including racism, xenophobia, and sexism) is an important avenue for future studies. 
We will also work on extending and improving NLP models for automatic aporophobia detection, with the ultimate goal of tracking and informing on the trends of aporophobia online over time.

This line of research aims to complement other lines of work on aporophobia, such as agent-based modeling simulations, that inform on the impact of poverty-mitigation policies under discussion at local parliamentary level \cite{aguilera2024can}. The capacity of NLP techniques to analyze and track complex social phenomena, such as aporophobia, over vast amounts of real-world data, combined with the capabilities of agent-based modeling to represent social impact contexts constitute a promising new set of tools to tackle poverty by acting on social biases.

\section*{Limitations}
As with any analysis of an intricate social challenge through computational techniques, this study has limitations. There are 2.6 billion people worldwide who do not have internet access \cite{InternationalTelecmmunicationsUnionITU2023}, and not all regions, genders, or identity groups are equally represented online \cite{AlanChan2021}. 
Among online users, we capture data only from users of the social media platform X, which is predominantly used in the United States \cite{barbieri-etal-2020-tweeteval} and
represents specific socioeconomic demographics in terms of age, gender, ethnicity, etc. \cite{Mislove2011}. Thus, many demographics are under-represented, and even if present, they might shy away from expressing their real beliefs. Further, the data is collected only in English and, mainly, the largest English-speaking countries are represented. In addition, the user posts are collected using a pre-specified set of terms in standard English that may exclude related terms in regional dialects. 
In general, data collection from social media may introduce different types of biases, such as selection bias, platform-induced behavior and public image curation bias. Therefore, the findings from this study may not generalize to population at large and should be considered preliminary. 

The data annotation has been performed by three annotators with diverse cultural backgrounds and familiarity with local events and societal expectations in different regions of the world. Still, misinterpretation of content and annotation bias may be present. 

Finally, it must be emphasized that, while allowing processing large amounts of data, NLP techniques can only analyze what people say, as opposed to what they think. Complementary data collection methods, such as the generation of an aporophobia-specific Harvard Implicit Association Test \cite{Xu2014}, workshops and interviews with the specialized non-profits and government officials, as well as surveys among the affected population will be considered to obtain additional quantitative and qualitative data. 

\section*{Ethics Statement}

Detecting, tracking, and mitigating the negative impact of aporophobia in the online social discourse poses a number of risks and ethical issues, 
discussed at length in previous works in the context of abusive and toxic language detection  
\cite{hovy-spruit-2016-social,vidgen-etal-2019-challenges,cortiz2020ethical,kiritchenko2021confronting}. 
These issues include tension between freedom of speech and respect for equality and dignity, biased data sampling and data annotation, dual use of technology, and many others.   
The research, design and deployment of such technology should comply with trustworthy AI principles of transparency, justice and fairness, non-maleficence, responsibility, and privacy \cite{Jobin2019}. 
This should include clear communication about the models' limitations, potential biases, and intended use cases. We would also like to emphasize the importance of involving key stakeholders---including non-profit organizations and underserved communities themselves---in the development process to ensure that the technology is aligned with broader social values and equipped with safeguards against potential harm.

\section*{Acknowledgments}

We would like to 
thank the teams in Arrels Fundacio, Caritas and Motels4Now, among many other contributors, for their inputs and guidance as domain experts. Their expertise has been key to this research and has allowed us to contextualize what aporophobia means in practice for the affected population.

\bibliography{references}

\appendix

\setcounter{table}{0}
\renewcommand\thetable{A.\arabic{table}}

\setcounter{figure}{0}
\renewcommand\thefigure{A.\arabic{figure}}

\section{Details on the Dataset Creation}

\subsection{Tweet Collection and Pre-processing}
\label{sec:appendix_tweets}

The initial query term list was constructed from social psychology literature and various web sources, and extended with synonyms from thesauruses. Then, terms resulted in few retrieved tweets or large amount of unrelated tweets in a preliminary search were removed.  
The single term \textit{poor} was excluded since it often appears in unrelated contexts (e.g., `poor taste', `poor results'). 
We also excluded derogatory terms, like `tramp', `beggar', or `trailer trash', since in our preliminary search we found these terms being disproportionately used as personal insults and not directly related to people living in poverty (for example, `\textit{You are a trailer trash queen who has no business in congress}').

Re-tweets, duplicate tweets, tweets with external URLs, and tweets with more than five hashtags were discarded. Further, we removed tweets originated from accounts with the word `bot' in their user name or screen name. 

For geographical location we used Twitter fields ‘place’ (tweet location) and `user location'. Only about 2\% of the collected tweets had exact tweet location (`place') specified. On the other hand, about 60\% of the tweets had user location field filled. The user location field is a free-form text, so we applied simple string matching to extract the most frequently mentioned countries. When country was not mentioned, we also tried to match U.S. states (or their two-letter abbreviations), Canadian provinces (or their two-letter abbreviations), and major cities in the U.S., U.K., and Canada. Since we collected tweets in English, most of the tweets originated from countries where English is commonly spoken. We grouped the tweets with known geographical location into the following regions: North America (U.S. and Canada), Europe (mostly U.K. and Ireland), Africa (Nigeria, South Africa, Kenya, Uganda, Ghana), South Asia (India, Pakistan, and Philippines), and Oceania (Australia and New Zealand). Tweets from other parts of the world or with unknown location were grouped as `Other'.

\begin{table*}
  \centering
  \small
  \begin{tabular}{p{0.5cm}p{6cm}p{8cm}}
    \hline
    \textbf{Topic} & \textbf{Topic words} & \textbf{Examples} \\
    \hline
    5 & drug, addicts, mental, drugs, mentally, ill, addiction, health, addicted, addict & Leave the drugs and alcohol out and you won’t be homeless, will you? \color{red}\textbf{[DIRECT]}\\
    6 & black, white, color, racist, minorities, blacks, whites, race, racism, brown & Most republicans think only Blacks are on welfare \color{purple}\textbf{[REPORTING]}\\
    10 & immigrants, migrants, illegals, illegal, asylum, refugees, seekers, hotels, border, immigration & We are overrun with refugees and not protecting our own homeless \color{red}\textbf{[DIRECT]}\\
    14 & crime, police, cops, criminals, jail, crimes, prison, arrest, commit, criminal & Poor people just steal other people's property instead of getting a job and earning money \color{red}\textbf{[DIRECT]}\\
    38 & hate, hating, hates, hatred, say, despise, just, hated, really, people & Tories just hate the poor \color{purple}\textbf{[REPORTING]}\\
    49 & stealing, steal, stole, stolen, thieves, theft, money, steals, thief, millions & It's disgusting to steal from the poor \color{teal}\textbf{[NONE]}\\
    56 & war, military, wars, army, soldiers, fight, join, recruitment, peace, die & Men in the military are from most disadvantaged families that have no other option \color{purple}\textbf{[REPORTING]}\\
    67 & crime, gas, prices, high, inflation, highest, taxes, higher, record, border & Too much crime, too much pollution, too much homeless \color{red}\textbf{[DIRECT]}\\
    88 & encampments, encampment, city, near, park, street, clear, looks, large, cleared & Clean the homeless encampments now! \color{red}\textbf{[DIRECT]}\\
    91 & smell, piss, bathroom, smells, toilet, bathrooms, toilets, smelling, restrooms, pee & Homeless swarm public transportation, forcing everyone to breathe in their body odor \color{red}\textbf{[DIRECT]}\\
    96 & justice, court, lawyers, lawyer, legal, courts, judges, judge, attorney, law & The objective for the creation of justice centres is to widen access of the poor to justice through legal aid \color{teal}\textbf{[NONE]}\\
    100 & drunk, beer, drink, drinking, alcohol, cigarette, drunks, drinks, liquor, smoking & I'd say, alcohol is the drink of the poor and frustrated, not the rich \color{red}\textbf{[DIRECT]}\\
    106 & law, laws, apply, rule, applies, rules, order, break, enforced, legal & Typical: one law for the rich, one law for the poor \color{purple}\textbf{[REPORTING]}\\
    118 & fear, scared, scary, anxiety, afraid, terrified, scare, terrifying, fears, mongering & The scariest thing in the world is to be homeless, right? \color{red}\textbf{[DIRECT]}\\
    139 & blame, blaming, fault, blamed, blames, problematic, problems, instead, mistakes, victim & Let's admit that we don't like the poor, blame them for their conditions, but have done nothing to help \color{purple}\textbf{[REPORTING]}\\
    \hline
  \end{tabular}
  \caption{Fifteen topics manually selected as the most relevant to aporophobia. The topics are ordered by the amount of tweets assigned to the topic. The topic words (from BERTopic) are the words that tend to appear frequently in the topic of interest, and less frequently in the other topics. Example tweets are manually selected as the most representative of the topic. Example tweets are paraphrased to preserve the anonymity of the users. The class label for each example is as annotated in the dataset.}
  \label{tab:topics}
\end{table*}

\subsection{Topic Modeling with BERTopic}
\label{sec:appendix_bertopic}

BERTopic \cite{grootendorst2022bertopic} is a flexible toolkit for supervised, semi-supervised, and unsupervised topic modeling. It utilizes a density-based clustering technique HDBSCAN \citep{campello2013density}, which produces clusters of arbitrary shapes and leaves some documents that do not fit any of the identified topics as outliers. 
The obtained topics/clusters are then represented with topic words, which are identified using class-based TF-IDF (c-TF-IDF). The `topic words' are defined as the words that tend to appear frequently in the topic of interest, and less frequently in the other topics. 

We ran BERTopic in the unsupervised mode with the following parameters. 
For converting text to numerical representations, we used the sentence transformers method based on the \textit{all-MiniLM-L6-v2} pre-trained embedding model.\footnote{\url{https://www.sbert.net/docs/pretrained_models.html}} 
For the vectorizer model, we used the CountVectorizer method,\footnote{\url{https://scikit-learn.org/stable/modules/generated/sklearn.feature_extraction.text.CountVectorizer.html}} and removed English stopwords and terms that appeared in less than 5\% of the sentences ($min\_df = 0.05$). 
For the HDBSCAN clustering algorithm, we specified the minimum size of the clusters as $min\_cluster\_size=500$. For all the other parameters, the default settings of the BERTopic package were used. 

BERTopic identified 142 topics and left about 42\% of tweets unclustered.  
We manually analyzed the topic words and the most representative example tweets from the obtained topics and selected 15 topics highly relevant to the concept of aporophobia. The selected topics, along with the topic words and  example tweets, are listed in Table~\ref{tab:topics}. 

\subsection{Human Annotators}

The three annotators of the DRAX dataset are authors of this paper. Two of them identify as females, and one as male. The ages vary from 20s to 40s. All three have received higher education at Western institutions, but have different cultural backgrounds. They have extensive knowledge on social biases, aporophobia, and NLP.

\subsection{Data Distribution}

Table~\ref{tab:data-stats-countries} shows the data distribution per country. Note that not all countries in a region are represented equally. For most regions, the overwhelming majority of tweets originate from the largest English-speaking country, such as the U.S. for North America and the U.K. for Europe.

\begin{figure}[t!]
  \includegraphics[width=0.95\columnwidth]{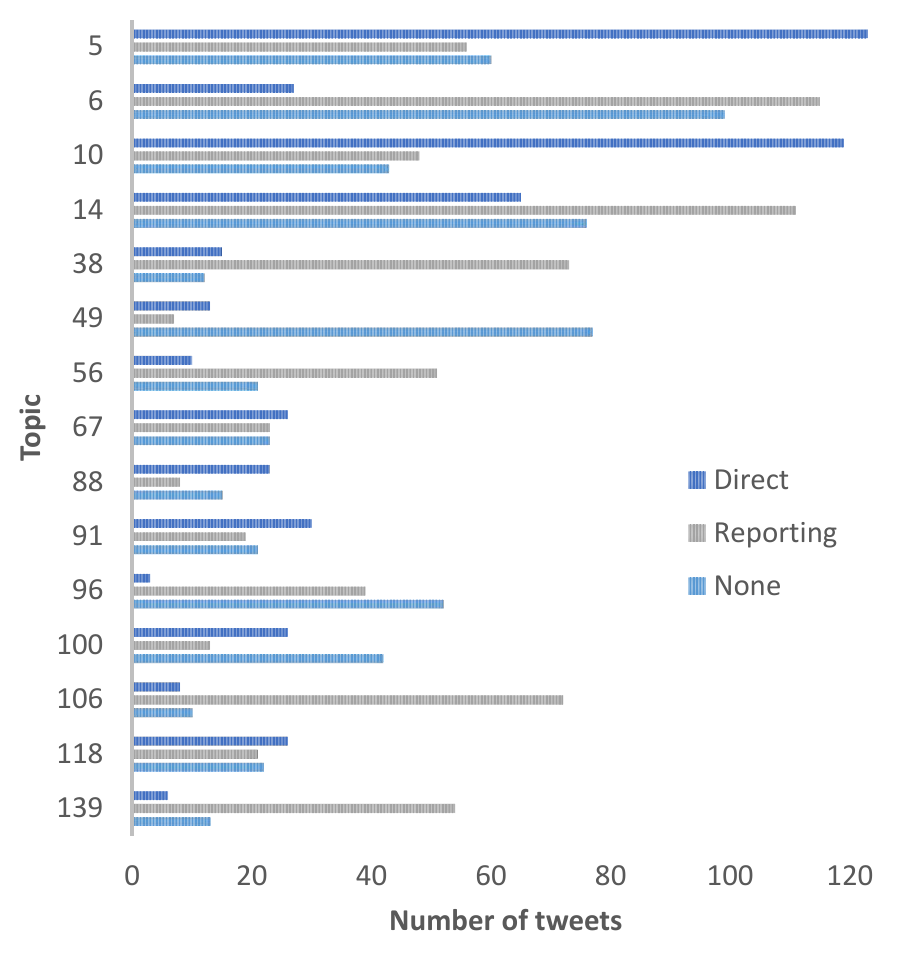}
  \caption{Class distribution per topic in the DRAX dataset.}
  \label{fig:data-stats-topics}
\end{figure}

Figure~\ref{fig:data-stats-topics} presents the class distribution per topic. Topics with the highest proportion of `Direct' aporophobic statements are 5 (drug addiction and mental health issues) and 10 (immigrants and refugees). Other  topics with a high proportion of the `Direct' category include 14 (crime), 67 (crime and other issues), 88 (homeless encampments), 91 (smell), 100 (alcohol addiction), and 118 (fear). 
Topics with a high rate of `Reporting' aporophobia are 6 (racism), 14 (crime), 38 (hatred), 56 (military), 96 (laws and courts), 106 (laws and regulations), and 139 (blame). 

\setlength{\tabcolsep}{3pt}
\begin{table}
  \centering
  \small
  \begin{tabular}{lrrrr}
    \hline
    \textbf{Country} & \textbf{Direct} & \textbf{Reporting} & \textbf{None} & \textbf{Total} \\
    \hline
    Ghana & 4 & 6 & 2 & 12\\
    Kenya & 7 & 22 & 11 & 40\\
    Nigeria & 17 & 26 & 27 & 70\\
    South Africa & 34 & 50 & 47 & 131 \\
    Uganda & 4 & 5 & 6 & 15\\
    \hline
    \textbf{Africa (total)} & \textbf{66}	& \textbf{109}	& \textbf{93} & \textbf{268}\\
    \hline
    \\
    France & 1 & 1 & 3 & 5\\
    Germany & 3 & 3 & 0 & 6\\
    Ireland & 9 & 17 & 6 & 32\\
    United Kingdom & 86 & 138 & 94 & 318\\
    \hline
    \textbf{Europe (total)} & \textbf{99}	& \textbf{159}	& \textbf{103}	& \textbf{361}\\
    \hline
    \\
    Canada & 5 & 16 & 8 & 29\\
    United States & 124 & 120 & 104 & 348\\
    \hline
    \textbf{North America (total)} & \textbf{129}	& \textbf{136}	& \textbf{112}	& \textbf{377}\\
    \hline
    \\
    Australia & 49 & 68 & 63 & 180\\
    New Zealand & 9 & 21 & 19 & 49\\
    \hline
    \textbf{Oceania (total)} & \textbf{58}	& \textbf{89}	& \textbf{82}	&\textbf{229}\\
    \hline
    \\
    India & 28 & 56 & 50 & 134\\
    Pakistan & 5 & 28 & 17 & 50\\
    Philippines & 3 & 10 & 5 & 18\\
    \hline
    \textbf{South Asia (total)} & \textbf{36}	& \textbf{94}	&\textbf{72}	&\textbf{202}\\
    \hline
    \\
    \textbf{Other (total)} & \textbf{132} & \textbf{136} & \textbf{111} & \textbf{379}\\
    \hline
  \end{tabular}
  \caption{The number of annotated tweets in the DRAX dataset per region and country.}
  \label{tab:data-stats-countries}
\end{table}
\setlength{\tabcolsep}{6pt}

\section{Additional Details and Experiments with NLP Models}

\subsection{Fine-tuned Language Models}
In this paper, we fine-tune several models from the BERT family to tackle the challenging task of aporophobia classification on social media. The first model, the original BERT (Bidirectional Encoder Representations from Transformers) \cite{BERT}, serves as the foundation for many modern NLP models. BERT, with its 110 million parameters, is well-suited for tasks like ours. By leveraging pre-training on large text corpora, BERT can be fine-tuned effectively for domain-specific applications. Next, we used RoBERTa (Robustly Optimized BERT Pretraining Approach) \cite{RoBERTa}, an advanced variant of BERT, which has the same architecture as BERT but uses a byte-level BPE as a tokenizer (same as GPT-2 \cite{GPT2}) and uses a different pretraining scheme. It has 125 million parameters and focuses solely on masked language modeling, which lead to improved performance in our classification experiments. We also fine-tune DistilBERT \cite{DistilBERT}, a distilled version of BERT designed to be more efficient without significantly compromising on accuracy. DistilBERT, with its reduced architecture of 66 million parameters, offers nearly the same language processing capabilities as BERT while being faster and more resource efficient, an important factor when deploying models at scale. Finally, we utilize BERTweet \cite{Bertweet}, a variant of BERT specifically pre-trained on English tweets. With the same architecture and 110 million parameters as BERT, BERTweet is particularly adept at handling the informal language found on social media platforms like Twitter. This specialization makes BERTweet an excellent candidate model for our task. 

The tweet pre-processing involved removing user mentions and URLs. 
The experiments were conducted using Google Colab Pro, leveraging their L4 GPU for faster training and inference. We utilized the models provided by the transformers library. The following hyperparameters were used for each model, representing the best-performing configurations after experimenting with different combinations:

\begin{itemize}
    \item \textbf{BERT, RoBERTa, DistilBERT:} Batch size = 4, Optimizer = Adam, Epochs = 4
    \item \textbf{BERTweet:} Batch size = 8, Optimizer = Adam, Epochs = 4
\end{itemize}
All reported results are the average of three independent runs with seed values of 42, 62, and 82 to ensure robustness and mitigate the impact of random initialization.

\subsection{Generative LLMs}
We used both open-source and closed-source models for our zero-shot and few-shot experiments with generative LLMs. For open-source models, we used Llama 3.1 405B Instruct, and Mixtral 8x22B Instruct. The Llama 3.1 405B is Meta's flagship language model with 405 billion parameters. It features a 128,000-token context window, enabling it to handle extensive textual inputs effectively. The Mixtral 8x22B is a Sparse Mixture-of-Experts (SMoE) model developed by Mistral AI, comprising 141 billion parameters, with only 39 billion active during inference. This architecture offers significant cost efficiency and performance advantages. The model supports a 64,000-token context window, facilitating precise information recall. 

In addition, we experimented with three commercial models via the OpenAI chat completions API\footnote{\url{https://platform.openai.com/docs/api-reference/chat/create}}: GPT-3.5 Turbo, GPT-4o mini, and GPT-4o.  
The GPT-3.5 Turbo 0125 model, with a 16,385-token context window, has been optimized for higher accuracy over GPT-3. GPT-4o mini, a smaller yet powerful model with a 128,000-token context window, has demonstrated more capabilities than GPT-3.5 Turbo and has shown efficiency on lightweight tasks. Finally, GPT-4o, the flagship OpenAI model with the same 128,000-token context window, was developed for handling more complex tasks. 

The following parameters were used for all models: temperature $t=0.7$, the maximum number of output tokens $max\_tokens=10$, and the number of chat completion choices $n = 1$.

Each of these models are tested to explore their performance in classifying aporophobic content with varying levels of prior information provided through prompts. The highest performance on DRAX was obtained with GPT-4o using a few-shot prompt listed in Table~\ref{tab:best-prompts}.

\setlength{\tabcolsep}{4pt}
\begin{table}
  \centering
  \small
  \begin{tabular}{lrrrr}
\hline
\textbf{Model}  & \textbf{Acc.} & \textbf{P} & \textbf{R} & \textbf{F1}   \\
\hline
ToxDect  & 0.63 & 0.59  & 0.63   & 0.60          \\
RoBERTa Toxicity& 0.67  & 0.63 & 0.67            & 0.64          \\
RoBERTa fine-tuned on DRAX & \textbf{0.79}     & \textbf{0.80}       & \textbf{0.79}   & \textbf{0.80}  \\

\hline
  \end{tabular}
\caption{Performance of two off-the-shelf RoBERTa-based toxicity models and the binary RoBERTa model fine-tuned on DRAX. The evaluation metrics are overall accuracy (Acc.), and support-weighted average precision (P), recall (R), and F1-score (F1).}
\label{tab:results-binary}
\end{table}
\setlength{\tabcolsep}{6pt}

\subsection{Experiments with Off-the-Shelf Toxicity Models}
\label{app:toxicity}

We experiment with two off-the-shelf toxicity detection models: ToxDect\footnote{\url{https://huggingface.co/Xuhui/ToxDect-roberta-large}} \cite{toxdect} and RoBERTa Toxicity Classifier\footnote{\url{https://huggingface.co/s-nlp/roberta_toxicity_classifier} This model is licensed under the OpenRAIL++ License.}  \cite{logacheva-etal-2022-paradetox}. Both models are based on the RoBERTa model and were fine-tuned specifically for toxicity classification tasks. We selected these models because they are publicly available on Huggingface, were trained on social media data, and are relatively recent. Additionally, their popularity on Huggingface, as indicated by a high number of downloads, further motivated their use, ensuring they are well-known within the community and suitable for our task. 

Given that both toxicity classifiers are binary classifiers returning `Toxic' or `Non-toxic' category labels, we adapt our dataset for a binary classification setting. Specifically, we merge the classes `Reporting' and `None' as `Non-toxic', and treat the `Direct' class as `Toxic'. For comparison, we also fine-tune a RoBERTa model on the DRAX training subset (modified for the binary classification setting). The results are shown in Table~\ref{tab:results-binary}. We observe that the RoBERTa model fine-tuned on the DRAX dataset substantially outperforms the pre-trained toxicity classifiers. This result indicates that while pre-trained toxicity models are generally good at identifying harmful speech, they are not optimized for the specific nuances of aporophobic content, highlighting the need for tailored models in this context. This conclusion is in line with the observations made by \citet{Kiritchenko2023} that existing toxicity and hate speech detection models and data resources are not effective for aporophobia detection.

\subsection{The Impact of Uniform Sampling across Regions}
\label{app:sampling}

To evaluate the impact of our oversampling strategy (i.e., uniform sampling from all six regional groups), we repeated the experiments with the training data sampled only from the dominant regions (North America and Other). We fine-tuned RoBERTa model on the training subset originated from North America and Other, and evaluate the performance of the model on the full test set. The results are shown in Table~\ref{tab:results-sampling}. Since the training set is now smaller than before (two regions vs.\@ six), the overall performance decreased. The largest losses in the weighted averaged F-score are observed for Oceania (11\%) and Africa (7\%), whereas the performance for North America decreased only by 3\%. Moreover, the results for the `Direct' category (the main category of interest) are the lowest for Africa and Oceania: F-score of 52\% and 53\% respectively, compared to 65\% for North America and 66\% for Other, with these differences driven predominantly by the differences in precision. These results demonstrate the importance of balanced data sampling across various geographical regions. 

\begin{table*}[h!]
  \centering
  \small
  \begin{tabular}{l|rrrr|r}
\hline
\textbf{Region}& \multicolumn{4}{c|}{\textbf{Training on North America \& Other}}  & \textbf{Full Training Set}  \\
& Direct & Reporting & None & Overall   & Overall   \\
\hline
Africa & 0.52 & 0.65 & 0.47 & 0.56 & 0.63\\
Europe & 0.56 & 0.67 & 0.45 & 0.57 & 0.61\\
North America & 0.65 & 0.68 & 0.58 & 0.64 & 0.67\\
Oceania & 0.53 & 0.74 & 0.38 & 0.57 & 0.68\\
South Asia & 0.62 & 0.76 & 0.45 & 0.63 & 0.67\\
Other & 0.66 & 0.61 & 0.32 & 0.55 & 0.61\\[5pt]
\hline
All & 0.61 & 0.68 & 0.45 & 0.58 & 0.64\\

\hline
  \end{tabular}
\caption{Performance of the RoBERTa-based model fine-tuned on the training data pertained to the dominant regions (North America and Other) and the RoBERTa-based model fine-tuned on the full training data. The evaluation metric is support-weighted average F1-score.}
\label{tab:results-sampling}
\end{table*}

\begin{table*}[h!]
\centering
\begin{tabular}{|p{15cm}|}
\hline
 \textbf{System prompt} \\ \hline
You are an AI trained to classify tweets based on their content into one of three categories: ``Direct'', ``Reporting'', or ``None'' with respect to aporophobia. \\ \hline
\textbf{User prompt} \\ \hline
Classify the tweet into one of these categories: ``Direct'', ``Reporting'' or ``None'' \\
Aporophobia involves negative sentiments and actions directed at the poor, including rejection, fear, and contempt. \\

\textbf{Direct:} The tweet shows the author's own aporophobic attitudes, such as derogatory remarks, negative stereotypes, or support for exclusionary practices. \\

\textbf{Reporting:} The tweet mentions or critiques aporophobic behaviors or attitudes of others without expressing the author's own negative views. \\

\textbf{None:} The tweet does not exhibit any aporophobic content or references. \\

\textbf{Examples:} \\

Tweet, Classification \\
``Let's be honest though, illegals are at least normal people trying to get a better life; homeless are usually mentally unwell junkies that would strip the copper out of the furniture for more crack'' , Direct \\
``No, we gonna get the streets cleaned up, stop crime and do something about the homeless invasion. NM is a shithole atm.'', Direct \\
``Put the homeless in jail and start work camps and chain gangs.'' , Direct \\
``Being poor does not take away your moral agency. Most poor people manage to not commit violent crime.'' , Reporting \\
``So a theft automatically equates to a homeless person?'' , Reporting \\
``Those kinda people never do time. Prison is for poor people.'' , Reporting \\
``Local charities are working hard to provide shelters and job training programs for the homeless to help improve their lives.'' , None \\
``Our city has launched a new initiative aiming to offer better healthcare access to the poor, ensuring everyone receives the care they need.'', None \\
``A recent study highlights the importance of community support in uplifting homeless individuals, demonstrating significant improvements in their well-being through local outreach programs.'' , None \\

Tweet: ``{Tweet}'' Limit the response to only one category from ``Direct'', ``Reporting'' or ``None'' \\

Select the category: \\ \hline
\end{tabular}
\caption{Best-performing few-shot prompt for GPT-4o.}
\label{tab:best-prompts}
\end{table*}

\newpage
\section{Annotation Guidelines}
\label{app:guidelines}

Aporophobia is defined as ``rejection, aversion, fear and contempt for the poor'' \cite{Cortina2022}. In order to annotate aporophobia in textual instances of social media, we make distinction between (1) direct instances of aporophobic beliefs and attitudes expressed by the speaker (toxic language) and (2) instances of aporophobic attitudes the speaker reports about others. Therefore, we have three main categories to be annotated: \textbf{`Direct Aporophobia’}, \textbf{`Reporting Aporophobia'}, and \textbf{`None’}. 

We note that aporophobia can manifest through the different degrees of action resulting from prejudice \cite{Allport1954}:
\begin{enumerate}
    \item \textbf{Antilocution or verbal rejection:} when an in-group freely purports negative images of an out-group (negative stereotypes, jokes, negative statements).
    \item \textbf{Avoidance and fear:} when members of the in-group actively avoid people in the out-group (expressing feelings of fear and the desire to avoid any contact).
    \item \textbf{Discrimination:} when the prejudiced person makes active detrimental distinctions, by denying the out-group opportunities and services. Note that segregation is considered an institutionalized form of discrimination, enforced either legally or by tradition. 
    \item \textbf{Physical attack:} Prejudices can also lead to acts of violence or semi-violence, such as forcibly evicting families from their homes or neighborhoods, or  physically attacking persons in a situation of homelessness. 
    \item \textbf{Extermination:} Lynchings, massacres are the ultimate degree of violent expression of prejudice. Even though we do not expect to see expressions of extermination directly linked to aporophobia on social media, we acknowledge that aporophobia can be an element in the prepared ground of previous hostility towards a particular group that is attacked and where other types of discrimination are intertwined. 
\end{enumerate}

Note that the categories `antilocution' and `avoidance and fear' can be expressed directly by the speaker on social media as well as reporting the attitudes of others, therefore, they can be labeled as either `Direct Aporophobia' or `Reporting Aporophobia'. In contrast,  the categories `discrimination', `physical attack' and `extermination', when they appear in textual instances, can only be considered `Reporting Aporophobia'. 

In reporting statements, we often find \textit{benevolent} instances where the speaker opposes the stereotype, such as in the following examples:  ``\textit{Most poor people manage to not commit violent crime.}'',\footnote{All example tweets in these guidelines are paraphrased to preserve the anonymity of the users.} ``\textit{I don't think you have to do drugs to be homeless}''. We consider these as instances of `Reporting Aporophobia' because, despite having a good intention, such messages indirectly acknowledge and reinforce the stereotype \cite{Beukeboom2019}.

These guidelines are based on our position that associating poor/homeless people with negative behavior, like addiction or crime, even if based on factual information, perpetuates the negative stereotypes and hinders social and economic measures to reduce poverty. We draw parallels with racism and sexism where stereotypical associations based on current standings (e.g., more doctors are male) are considered not acceptable for a just society.

\vspace{5mm}
\noindent\textbf{DIRECT APOROPHOBIA}
\vspace{2mm}

\noindent\textbf{Antilocution:} includes, but is not limited to the following topics: 
\begin{itemize}
    \item Associating poverty with laziness and taking advantage of public resources: ``\textit{Homeless population don’t pay taxes and get lots of freebees}''; ``\textit{Poor people always blame others for their misfortunes.}''; ``\textit{Rich people see opportunities and focus on rewards. Poor people see obstacles and focus on the risks}''; ``\textit{Most poor people don't want to try new careers or new business opportunities}''.
    \item Associating the poor with addiction: ``\textit{Do you realize that most of these folks are homeless BECAUSE they’re addicts and not the other way around?}''; ``\textit{If you don’t subsidize drug users and the homeless issue will get cut in half overnight}'' 
    \item Associating the poor with mental illness: ``\textit{Most homeless are mentally ill, so put them into a home for the mentally ill. If there is a shortage of institutional homes, build more. Problem solved.}''; ``\textit{Let’s be honest, illegals are at least normal people, homeless are usually mentally unwell junkies that would do anything for more crack.}'' 
    \item Associating the poor with crime, which can be divided into: 
    \begin{itemize}
    \item Overestimating the correlation between poverty and crime: ``\textit{Most of the students of such institutions are from poor families who are even willing to commit state organized crimes}''; ``\textit{But crime is a route that many poor people take just to survive.}''
    \item The criminalization of poverty: ``\textit{put the homeless in jail and start work camps}''; ``\textit{ENFORCE THE LAW ON THESE CRIMINAL ``HOMELESS'' who refuse services.}''; ``\textit{Prison would be a lot more comfortable than cold pavements and begging for cash.}''
\end{itemize}
    \item Associating poverty with bad hygiene: ``\textit{you can stink like the poor}''; ``\textit{smells like poor people in here}''; ``\textit{I can’t get the smell of Poor People out of my hands}''
\end{itemize}

\noindent\textbf{Avoidance and fear:} includes, but is not limited to the following topics: 
\begin{itemize}
    \item Exclusion, detachment and ostracizing:  ``\textit{We gonna get the streets cleaned up and end the homeless invasion}''; ``\textit{That is the place to move!! No homeless and no migrants!!}''
    \item Fear of poor / homeless people: ``\textit{I check every room in my house for a homeless man who could be lurking in my carpet}''; ``\textit{homeless-phobia is a big reason things have changed.}''
    \item Fear of being poor / homeless: ``\textit{Being homeless is the scariest thing ever.}'' Note: expressing concern about being in a situation of poverty is understandable since one might not have the necessary resources to cover one's needs and conduct a meaningful life with dignity. We consider specific expressions in this category when we believe they also imply social stigma or ostracism. 
\end{itemize}

\vspace{5mm}
\noindent\textbf{REPORTING APOROPHOBIA}
\vspace{2mm}

\noindent\textbf{Antilocution:} includes, but is not limited to the following topics: 
\begin{itemize}
    \item Associating poverty with laziness and taking advantage of public resources: ``\textit{Me or anyone on social welfare is not a burden. Our situations are not unique, this could happen to anyone.}''; ````\textit{Should we throw more taxpayers money at them?'' You seem to be blaming the entire `disadvantaged' community}''; ``\textit{it’s the `being homeless is their own fault and I should be allowed to shoot them with a gun' part that turns people off.}''
    \item Associating the poor with addiction: ``\textit{not everyone is homeless due to addiction}''; ``\textit{the guy uses that hateful term `open air drug markets' to refer to homeless encampments}''; `\textit{It's wrong to assume people become homeless due to drugs, I've never touched drugs in my life, but was homeless for a year after my divorce.}''
    \item Associating the poor with mental illness: ``\textit{This man is talking about how homeless people should just all be thrown into insane asylums.}'' 
    \item Associating the poor with crime, which can be divided into:  
\begin{itemize}
    \item Overestimating the correlation between poverty and crime: ``\textit{So a theft automatically equates to a homeless person?}''
    \item The criminalization of poverty: ``\textit{if it were up to you, you’d fill up every jail with the homeless}''; ``\textit{For them, the law is for the perpetual enslavement of the poor}''; ``\textit{The reality is that American law and order is brutal on the poor.}''
\end{itemize}
    \item Associating poverty with bad hygiene: ``\textit{If a homeless person shits on the street because they can't access a bathroom, it's an excuse to evict and attempt to exterminate}''; ``\textit{Unless you want people to think the poor are dirty and smelly...which they are not!}''
\end{itemize}

\noindent\textbf{Avoidance and fear:} includes, but is not limited to the following topics: 
\begin{itemize}
    \item Exclusion, detachment and ostracizing:  ``\textit{The obsession with trying to ``improve'' the lives of the disadvantaged, without those being discussed having a say.}''; ``\textit{Humane society takes dog from homeless person arguing it deserves a better life and leaves person on street}''; ``\textit{No one left behind” Except the poor, the disabled and the asylum seekers}''
    \item Fear of poor / homeless people: ``\textit{Nah. It’s just political opportunism. ABC has staked their entire campaign on fear of poor people}''; ``\textit{Hating homeless addicts is literally based on a childish fear of them}''; ``\textit{There is no crime or homeless problem. There is only fear-mongering.}''; ``\textit{If you find homeless people `scary and uncomfortable', give them homes.}'' 
\end{itemize}

\noindent\textbf{Discrimination:} includes, but is not limited to the following topics: 
\begin{itemize}
    \item Bullying:  ``\textit{Blaming the poor is a great tactic to get people on your side.}''; ``\textit{They want to grind homeless addicts under the boot or ship them somewhere else}''  
    \item Over-policing and criminalization:  ``\textit{All he does is harass homeless people and the local chapter/other leftist orgs}''; ``\textit{it is a horribly broken system of police training environment that prioritizes policing the poor, leading to brutality and overreactions}''
    \item Law / regulation enforcement: ``\textit{Those kinds of people never do time. Prison is for poor people.}''; ``\textit{It is more profitable to try, convict, and incarcerate poor people}''; ``\textit{Too many poor people in jail for minor offenses and it has nothing to do with community safety.}''  
    \item Military / war service:  ``\textit{A lot of poor people joined the military in the day as Black and White ghetto kids often had poor nutrition at home}''; ``\textit{They are actually admitting that the military depends on recruiting poor people.}''
\end{itemize}

\noindent\textbf{Physical attack:} includes examples such as ``\textit{Police charged for pouring oil on the homeless}''; ``\textit{Two arrested for beating and looting a homeless old man around midnight 2 days back.}''; ``\textit{That is a clever little plan to pay young people to beat up homeless people'}'; ``\textit{Maybe concentrate on stopping the homeless being abused on the streets.}'' 

\vspace{5mm}
\noindent\textbf{Bias Aggravation}
\vspace{2mm}

It is important to highlight that aporophobia can act as an aggravator of other types of discrimination (namely, racism, xenophobia, and sexism) and in many instances these different types of discrimination appear intertwined. For example, a recurrent argument that appears on social media is the ``need to take care of our homeless first'' as an argument to reject migrants and ethnic minority groups. We consider the following examples as `Direct Aporophobia' because, in this case, poverty is part of the argument to reject ethnic minority groups and migrants (i.e., if these population groups were rich, they would not be rejected with the argument that they are competing for resources with the local population). In other words, when the minority group is poor, this contributes to racism or xenophobia. Some examples are as follows: ``\textit{We don't want them! Can't help our homeless, but sure let's put a roof over migrants heads. No thanks, send them back where they came from or dump them elsewhere.}''; ``\textit{They spend millions to house, feed and educate immigrants all across America in ways which they do not do for the Homeless Americans in their own cities}''; ``\textit{BLACK people are more likely to be homeless and need financial support.}''

\end{document}